\documentstyle[11pt]{article}

\begin{document}
\newcommand{\BE}{\begin{equation}}
\newcommand{\EE}{\end{equation}}
\newcommand{\R}{R_{e^+e^-}}
\newcommand{\lam}{\tilde{\Lambda}}
\begin{titlepage}
\begin{flushright}
{\small DE-FG05-92ER40717-2}
\end{flushright}
\vspace*{34mm}
\begin{center}
{\Large \bf Response to Brodsky and Lu's Letter \\
\vspace*{2mm}
          ``On the Self-Consistency of Scale-Setting Methods''}
\vspace{29mm}\\
{\Large P. M. Stevenson}
\vspace{18mm}\\
{\large\it
T.W. Bonner Laboratory, Physics Department,\\
Rice University, Houston, TX 77251, USA}
\vspace{30mm}\\
{\bf Abstract:}
\end{center}

   The claim that the Principle of Minimal Sensitivity is ``disfavored''
because it does not satisfy certain ``self-consistency requirements''
is meaningless, and shows a basic misunderstanding of the
renormalization-scheme-dependence problem.

\end{titlepage}
\setcounter{page}{1}

   Brodsky and Lu (BL) \cite{BL} have criticized the ``Principle of
Minimal Sensitivity'' (PMS) optimization method \cite{OPT}, claiming that
it fails to satisfy certain ``reflexivity'', ``symmetry'', and ``transitivity''
conditions.  These conditions involve the series expansion of one couplant
($a = \alpha_s/\pi$) in terms of another:
\BE
\label{eq1}
a_1(\mu_1) = a_2(\mu_2)(1 + r_{12}(\mu_1,\mu_2) a_2(\mu_2) + \ldots).
\EE
BL apply PMS to this series as though the left-hand side were a physical
quantity.  This is a basic error: couplants (and Green's functions,
anomalous dimensions, etc.) are not physical quantities and are not
renormalization-group (RG) invariant.  PMS relies on the knowledge that the
quantity in question is RG invariant -- or would be, were it not for the
truncation of the perturbation series.  It makes no sense to apply PMS to
``optimize'' $a_1(\mu_1)$ because it is not RG invariant.  Moreover,
there is no {\it point} in trying to optimize things that are not physical
quantities; they are ill-defined (being intrinsically scheme dependent), and
they cannot be compared to experiment, anyhow.  Equations like (1) contain no
information about physics.  Thus, BL's conditions are meaningless.

   [It is quite amusing, though, to note that if optimization of (1) and
``self-consistency'' arguments were meaningful, then they would favour
PMS over other schemes.  Suppose we take a second-order approximation to
(1):
\BE
\label{trunc}
a_1^{(2)} \equiv a_2^{(2)} (1 + r_{12} a_2^{(2)} ),
\EE
where $a_2^{(2)}$, by definition, satisfies the second-order truncated
$\beta$-function:
\BE
\label{beta2}
\mu_2 \frac{\partial a_2^{(2)}}{\partial \mu_2} = - b (a_2^{(2)})^2
(1 + c a_2^{(2)}).
\EE
Now, consider the $\mu_1$ dependence of $a_1^{(2)}$:
\BE
\mu_1 \frac{\partial a_1^{(2)}}{\partial \mu_1} =
\mu_1 \frac{\partial r_{12}}{\partial \mu_1} (a_2^{(2)})^2 = - b (a_2^{(2)})^2.
\EE
Substituting for $a_2^{(2)}$ in terms of $a_1^{(2)}$ from (\ref{trunc})
(reversed) yields:
\BE
\mu_1 \frac{\partial a_1^{(2)}}{\partial \mu_1} = - b \left[
a_1^{(2)} ( 1 - r_{12} a_1^{(2)} + \ldots) \right]^2
\nonumber
\EE
\BE
\mbox{\hspace*{1cm}}
= - b (a_1^{(2)})^2 ( 1 - 2 r_{12} a_1^{(2)} + \ldots).
\EE
But, for ``self-consistency'' in BL's sense, we would want $a_1^{(2)}$ to
satisfy the second-order $\beta$-function equation (i.e., Eq. (\ref{beta2})
with `2' subscripts replaced by `1' subscripts), at least up to higher-order
corrections.  In FAC, where the ``optimization'' gives
$\overline{r_{12}}({\scriptstyle{FAC}}) = 0$, this would not be true.  Only in
PMS, where the optimization gives
$\overline{r_{12}}({\scriptstyle{PMS}}) = - \frac{1}{2} c + {\cal O}(\bar{a})$
is this ``self-consistency'' condition satisfied.  I don't take this very
seriously, since the overriding point is that Eq. (1) contains no physics,
but it does show that ``self-consistency'' arguments are a double-edged sword.]

   BL also incorrectly state that ``there are no known theorems that
guarantee the existence or the uniqueness of the PMS solution''.  In second
order it is trivial to prove this from the explicit functional form of
the second-order approximant ${\cal R}^{(2)}$ considered as a function of
$a^{(2)}$, the second-order couplant, which is itself a function only of the
scheme variable $\mu/\lam$ \cite{OPT}.  Completely generally it can be stated
that ${\cal R}^{(2)}$ rises from zero at $a=0$, has a single maximum, and then
becomes negative.  There may be spurious stationary points beyond this, or at
negative $a^{(2)}$, but these are obviously irrelevant.  There is always one
and only one solution in the physically relevant region.  Furthermore, a
simple all-orders argument can be given (see Ref. \cite{OPT}, Sect. V.D,
remark (1)).  One can set up a well-defined procedure for obtaining the
$n$th-order PMS-optimized approximant as a series of definite higher-order
correction terms to the initial $n$th-order approximant in some initial,
arbitrarily chosen RS.  The correction terms in this ``improvement
formula'' are uniquely determined.\footnote{
Another, better, way of obtaining an approximate solution to the PMS
optimization equations is the PWMR approximation \cite{PWMR}.  This can
also be systematically improved in an unambiguous way.}
This indicates that in all orders one will find a unique, perturbatively
relevant, solution to the PMS optimization equations.

   BL state their first condition in terms of the ``existence and
uniqueness of $\mu$'', which betrays another basic misunderstanding -- one
that is, alas, all too common.  My point is that the relevant variable
(and in second order the {\it only} relevant variable) is not $\mu$ but
the {\it ratio} $\mu/\lam$ \cite{OPT,sn}.  In PMS (or in FAC, or in any
method that is not pure nonsense) there is no such thing as an optimal scale,
$\bar{\mu}$.  There {\it is} an optimal value of the ratio $\mu/\lam$, and
this is what matters.  For a physical quantity ${\cal R}$:
\BE
\label{R}
{\cal{R}} = a(\mu)^p(1+r_1 a(\mu) + \ldots ),
\EE
the coefficient $r_1$ is dependent both on the scale $\mu$ and on the
renormalization convention used in defining the couplant.  However, these
dependences both arise from $\mu/\lam$ dependence, and one can prove
\cite{OPT} that the combination:
\BE
\rho_1 = b \ln (\mu/\lam) - r_1/p
\EE
(where $b$ is the leading $\beta$-function coefficient), is RS-invariant.
That is, not only does the $\mu$ dependence cancel, but the convention
dependence of $\lam$ cancels the convention dependence of $r_1$.
To prove this one needs the Celmaster-Gonsalves relation \cite{CG}
(see also Appendix A of Ref. \cite{OPT}), which states that in two conventions
related by
\BE
a'(\mu) = a(\mu) ( 1 + v_1 a(\mu) + v_2 a(\mu)^2 + \ldots),
\EE
the two $\lam$ parameters are related {\it exactly} by:
\BE
\ln (\lam'/\lam) = v_1/b.
\EE

  The point that $\mu/\lam$ is what matters -- not scale or convention
dependence separately -- is crucial to an understanding of the RS-dependence
problem \cite{sn}.  Methods that purport to deal with the RS-ambiguity problem
by proposing a unique renormalization convention -- leaving the scale to be
fixed independently -- are useless because they don't fix $\mu/\lam$, the
variable that embodies the ambiguity.  This criticism applies with equal force
to ``scale-fixing'' methods that purport to determine an optimal scale
$\bar{\mu}$ independent of the renormalization-convention choice.  The
so-called BLM method, as was pointed out very gently long ago \cite{CS}, is
useless for just this reason.

  In another paper \cite{BL2} BL claim to have devised a formalism that
avoids RS ambiguities altogether.  This is fallacious, as are all such claims
to have invented ``RS-invariant perturbation theory''.  The RS ambiguity
inevitably enters because at some stage one must make a choice of expansion
parameter, $a$, and in Ref. \cite{BL2} the arbitrary choice $a={\cal R}$ is
made.  This is simply the ``fastest apparent convergence'' (FAC) scheme
\cite{grun}.  For a critical discussion of the arbitrariness of the
$a={\cal R}$ scheme choice, see Ref. \cite{chyla}.  In practical QCD
applications FAC leads to results quite close to PMS, but in other contexts
FAC can fail quite badly \cite{sn}.

\vspace*{4mm}
\hspace*{-\parindent}{\bf Acknowledgements}

This work was supported in part by the U.S. Department of Energy under
Grant No. DE-FG05-92ER40717.

\end{document}